\documentclass[a4paper]{jpconf}
\usepackage{graphicx}
\begin{document}
\title{Sharp Valence Change as Origin of Drastic Change of Fermi Surface and Transport Anomalies 
in CeRhIn$_5$ under Pressure}

\author{Shinji Watanabe, Kazumasa Miyake}

\address{Graduate School of Engineering Science, Osaka University, Toyonaka, Osaka 560-8531, Japan}

%\ead{}

\begin{abstract}
The drastic changes of Fermi surfaces as well as transport anomalies near $P=P_{\rm c}\sim 2.35$ GPa in CeRhIn$_5$ are explained theoretically from the viewpoint of sharp valence change of Ce. It is pointed out that the key mechanism is the interplay of magnetic order and Ce-valence fluctuations. It is shown that the antiferromagnetic state with "small" Fermi surfaces changes to the paramagnetic state with "large" Fermi surfaces with huge enhancement of effective mass of electrons with keeping finite c-f hybridization. This naturally explains the de Haas-van Alphen measurement and also the transport anomalies of $T$-linear resistivity emerging simultaneously with the residual resistivity peak at $P=P_{\rm c}$ in CeRhIn$_5$.
\end{abstract}

\section{Introduction}

A heavy electron metal CeRhIn$_5$ has attracted much attention recently since it has been recognized 
to offer a new insight into the quantum critical phenomena by accumulated experimental studies. 
By applying pressure under the magnetic field larger than the upper critical field, 
the antiferromagnetic (AF) order is suppressed abruptly at $P=P_{\rm c}\approx 2.35$~GPa~\cite{Shishido2005}. 
The de Haas-van Alphen (dHvA) measurement revealed that the Fermi surfaces (FS) change from approximately 
the same FS as those of LaRhIn$_5$ to FS similar to those of CeCoIn$_5$ at $P=P_{\rm c}$ with huge mass enhancement 
toward $P=P_{\rm c}$. 
Transport anomalies such as emergence of $T$-linear resistivity 
and enhancement of residual resistivity are prominent near $P=P_{\rm c}$~\cite{Park2006,Park2008,Knebel2008}. 
So far, it has been considered that localized-to-itinerant transition of f electrons 
might explain these anomalies~\cite{Park2008}. 
However, this scenario encounters a serious difficulty in elucidating the experimental fact 
that the Sommerfeld constant at ambient pressure $P=0$ in CeRhIn$_5$, $\gamma=50$~mJ/mol{K$^2$}, 
is about 10-times larger than that of LaRhIn$_5$~\cite{Shishido2005}, 
which strongly suggests that heavy quasiparticles are formed in the AF-ordered state, 
showing an importance of hybridization between f and conduction electrons. 

In this paper, we propose a unified explanation for resolving this outstanding puzzle in CeRhIn$_5$. 
First, we point out that the emergence of $T$-linear resistivity and residual-resistivity peak 
are quite similar to the observations in CeCu$_2$Ge$_2$~\cite{jaccard}, 
CeCu$_2$Si$_2$~\cite{holms}, and CeCu$_2$(Si$_{x}$Ge$_{1-x}$)$_2$~\cite{yuan} 
at pressure $P=P_{\rm v}$ located at higher pressure side than the AF-paramagnetic boundary $P=P_{\rm c}$. 
The Cu-NQR measurement detected that the electric-field gradient starts to change at $P=4$~GPa 
slightly less than $P_{\rm v}\approx 5$~GPa where Ce valence changes~\cite{fujiwara}. 
Theoretically, it has been shown that enhanced valence fluctuations cause 
the $T$-linear resistivity~\cite{holms} and residual-resistivity peak at $P=P_{\rm v}$~\cite{MM}. 
Since in CeRhIn$_5$ such a transport anomalies occur at the pressure of 
the AF-paramagnetic boundary, $P_{\rm c}$ seems to almost coincide with $P_{\rm v}$, i.e., 
$P_{\rm c}\approx P_{\rm v}$. 
Below we demonstrate theoretically that this situation is indeed realized for realistic parameters 
for CeRhIn$_5$ and show 
that the interplay of the AF order and Ce-valence fluctuations is a key mechanism for resolving 
the outstanding puzzle in CeRhIn$_5$ in a unified way~\cite{CeRhIn5}.

\section{Model Calculation and Results}

Let us start our discussion by introducing a minimal model which describes 
the essential part of the physics of CeRhIn$_5$ in the standard notation: 
\begin{equation}
{\cal H}=H_{\rm c}+H_{\rm f}+H_{\rm hyb}+H_{U_{\rm fc}}, 
\label{eq:PAM} 
\end{equation}
where 
$H_{\rm c}=\sum_{{\bf k}\sigma}\varepsilon_{\bf k}
c_{{\bf k}\sigma}^{\dagger}c_{{\bf k}\sigma}$ 
represents the conduction band, 
$H_{\rm f}=\varepsilon_{ \rm f}\sum_{i\sigma}n^{ \rm f}_{i\sigma}
+U\sum_{i=1}^{N}n_{i\uparrow}^{ \rm f}n_{i\downarrow}^{ \rm f}$ 
the f level $\varepsilon_{\rm f}$ and onsite Coulomb repulsion $U$ for f electrons, 
$H_{\rm hyb}=V\sum_{i\sigma}\left(
f_{i\sigma}^{\dagger}c_{i\sigma}+c_{i\sigma}^{\dagger}f_{i\sigma}
\right)$ 
the hybridization $V$ between f and conduction electrons, 
and 
$
H_{U_{\rm fc}}=
U_{\rm fc}\sum_{i=1}^{N}n_{i}^{ \rm f}n_{i}^{ c}
$ 
the Coulomb repulsion $U_{\rm fc}$ between f and conduction electrons. 
The $H_{U_{\rm fc}}$ term is a key ingredient for explaining the various anomalies 
caused by enhanced Ce-valence fluctuations: 
The $T$-linear resistivity and residual resistivity peak have been shown theoretically 
on the basis of this model eq.~(\ref{eq:PAM})~\cite{holms,MM}. 
To discuss CeRhIn$_5$, we analyze this model by taking into 
account existence of the AF order. 
To make a comparison with the dHvA measurement 
for the $\beta_2$ branch which is a two-dimensional-like cylindrical Fermi surface, we consider 
the AF order with the ordered vector ${\bf Q}=(\pi,\pi)$ on the square lattice 
with $\varepsilon_{\bf k}=-2t(\cos(k_x)+\cos(k_y))$ 
at the filling $n\equiv (\bar{n}_{\rm f}+\bar{n}_{\rm c})/2=0.9$ 
with $\bar{n}_{\rm f}\equiv\sum_{{\bf k}\sigma}\langle 
f^{\dagger}_{{\bf k}\sigma}f_{{\bf k}\sigma}
\rangle/N$ 
and 
$\bar{n}_{\rm c}\equiv\sum_{{\bf k}\sigma}\langle 
c^{\dagger}_{{\bf k}\sigma}c_{{\bf k}\sigma}
\rangle/N$. 
We take $t=1$ as an energy unit. 

\begin{figure}[t]
\includegraphics[width=150mm]{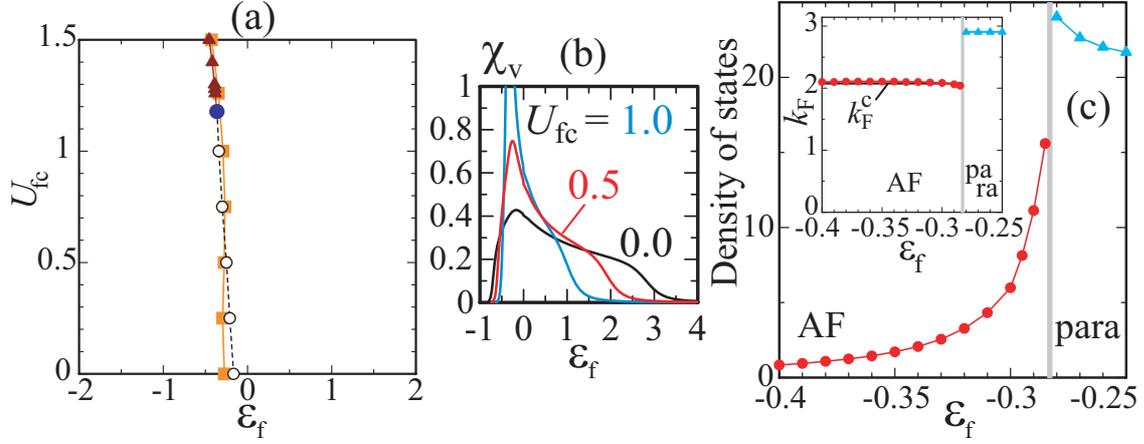}
\caption{\label{fig:PD}(color online) 
(a) Ground-state phase diagram in the $U_{\rm fc}$-$\varepsilon_{\rm f}$ plane 
for paramagnetic and AF states (see text). 
The first-order valence-transition line (solid line with triangles) 
terminates at the quantum critical end point (filled circle). Valence crossover 
occurs at the dashed line with open circles, at which $\chi_{\rm v}$ has a maximum, 
as shown in (b). 
The solid line with filled squares represents the boundary between the AF state 
and the paramagnetic state. 
(c) Total density of states vs $\varepsilon_{\rm f}$ for $U_{\rm fc}=0.5$. 
The inset shows $k_{\rm F}$ vs $\varepsilon_{\rm f}$: 
In AF phase, 
black line represents the small Fermi surface of conduction electrons 
$\varepsilon_{\bf k}$ at $\bar{n}_{\rm c}=0.8$. 
All results in (a)-(c) are calculated for 
$t=1$, $V=0.2$, and $U=\infty$ at $n=0.9$. 
}
\end{figure}

To treat the effects of the AF order and Ce-valence fluctuations 
on equal footing, we apply the slave-boson mean-field theory~\cite{KR} to 
eq.~(\ref{eq:PAM})~\cite{CeRhIn5}. 
The resultant ground-state phase diagram is shown in Fig.~\ref{fig:PD}(a). 
The first-order valence transition (solid line with triangles) terminates at the 
quantum critical end point (filled circle), i.e., the QCP, and the valence crossover occurs 
at the dashed line with open circles. 
At the QCP the valence fluctuation diverges and 
the enhanced valence fluctuations appear even far away from the QCP, 
as indicated by enhanced valence susceptibility 
$\chi_{\rm v}\equiv-\partial\bar{n}_{\rm f}/\partial\varepsilon_{\rm f}$ 
in Fig.~\ref{fig:PD}(b) 
calculated for moderate $U_{\rm fc}$'s. 
These are the results obtained within the paramagnetic solutions. 
The result when the AF order is taken into account is shown as the 
solid line with filled squares in Fig.~\ref{fig:PD}(a). 
One can see that the AF-paramagnetic boundary, which is the first-order transition, almost coincides with the 
first-order valence-transition line and the valence-crossover line. 
This result is favorably compared with the observations in CeRhIn$_5$. 
When pressure is applied to the Ce compounds, $\varepsilon_{\rm f}$ increases. 
Thus, $\varepsilon_{\rm f}$ can be regarded as pressure. 
As shown in Fig.~\ref{fig:PD}(a), the AF order is suppressed at the Ce-valence 
transition and crossover points, which indicates that the suppression is caused by 
the valence transition or enhanced valence fluctuations.  
This result implies that $P_{\rm c}\approx P_{\rm v}$ observed in CeRhIn$_5$ is 
reproduced by our model eq.~(\ref{eq:PAM}). 
For comparison, hereafter, we show our results 
for $U_{\rm fc}=0.5$, which is a  moderate $U_{\rm fc}$ 
as expected to be a realistic parameter for CeRhIn$_5$.

%%%%%%%%%%%%%%%%%%%%%%  Fig.1 %%%%%%%%%%%%%%%%%%%%%%%%%%%%%%%%%%%%%%%%%%%%%%%%%%%%%%%%%%%%%%%%%%
%%%%%%%%%%%%%%%%%%%%%%%%%%%%%%%%%%%%%%%%%%%%%%%%%%%%%%%%%%%%%%%%%%%%%%%%%%%%%%%%%%%%%%%%%%%%%%%%

\begin{figure}
\includegraphics[width=160mm]{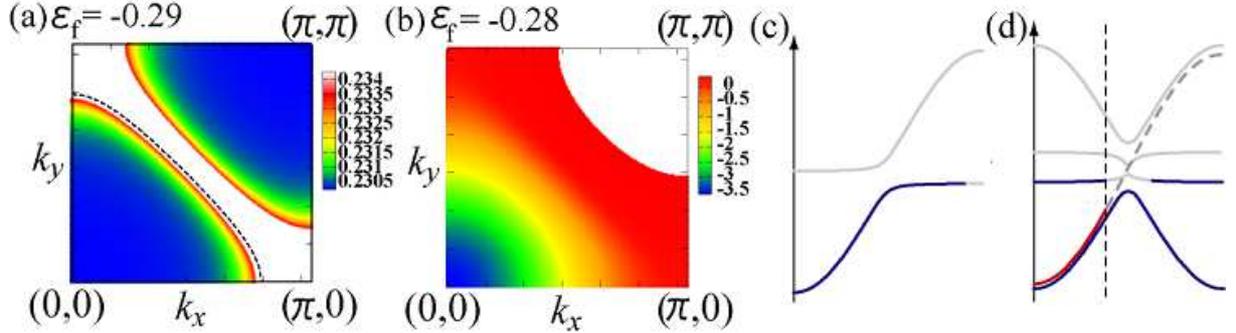}
\caption{\label{fig:FS}(color online) 
The contour plot of the energy band with $\downarrow$ spin 
located at the Fermi level $\mu$ 
for $t=1$, $V=0.2$, $U=\infty$, $U_{\rm fc}=0.5$, and $n=0.9$ with $h=0.005$: 
(a) $\varepsilon_{\rm f}=-0.29$ and 
(b) $\varepsilon_{\rm f}=-0.28$. 
The $E_{{\bf k}\downarrow}>\mu$ parts are represented by white regions. 
In (a), the dashed line indicates the Fermi surface of the conduction band, 
$\varepsilon_{\bf k}$ for $\bar{n}_{\rm c}=0.8$. 
Occupied (solid lines) and empty (gray lines) 
bands in the periodic Anderson model at $n=(\bar{n}_{\rm f}+\bar{n}_{\rm c})/2=0.9$ are shown 
in paramagnetic phase (c) and in AF-ordered phase (d). 
In (d), gray dashed line indicates the energy band of the conduction band, 
$\varepsilon_{\bf k}$ at $\bar{n}_{\rm c}=0.8$. 
Solid dashed line is a guide for the eyes indicating that the Fermi surface in AF-ordered phase 
with finite c-f hybridization coincides with the small Fermi surface where f electrons are 
completely localized. 
}
\end{figure}

Figures~\ref{fig:FS}(a) and \ref{fig:FS}(b) show the contour plots of the lower hybridized band with minority spin for $\varepsilon_{\rm f}=-0.29$ and $\varepsilon_{\rm f}=-0.28$, 
respectively. Here, we apply the magnetic field to the system as ${\cal H}-h\sum_{i}(S_{i}^{{\rm f}z}+S_{i}^{{\rm c}z})$ 
with $h=0.005$ to simulate the field of $H=15$~T used in the dHvA measurement~\cite{Shishido2005}. 
%The AF-to-paramagnetic transition occurs at $\varepsilon_{\rm f}^{\rm c}=-0.283$. 
We have verified that the drastic change of the Fermi surface occurs at 
the AF-paramagnetic-transition point 
$\varepsilon_{\rm f}=\varepsilon_{\rm f}^{\rm c}=-0.283$: 
Folded Fermi surface in Fig.~\ref{fig:FS}(a) in the AF phase changes to the unfolded Fermi surface in Fig.~\ref{fig:FS}(b) 
in the paramagnetic phase as $\varepsilon_{\rm f}$ passes across $\varepsilon_{\rm f}^{\rm c}$.
We note here that the dashed line represents the Fermi surface of conduction electrons 
at $\bar{n}_{\rm c}=0.8$. 
This is the small Fermi surface obtained by setting $V=0$ in the model eq.(\ref{eq:PAM}), 
where f electrons with $\bar{n}_{\rm f}=1$ are completely localized. 
One can see that the Fermi surface in the AF phase calculated for finite $V$ is nearly the same 
as the ``small" Fermi surface.

%%%%%%%%%%%%%%%%%%%%%%  Fig.2 %%%%%%%%%%%%%%%%%%%%%%%%%%%%%%%%%%%%%%%%%%%%%%%%%%%%%%%%%%%%%%%%%%
%%%%%%%%%%%%%%%%%%%%%%%%%%%%%%%%%%%%%%%%%%%%%%%%%%%%%%%%%%%%%%%%%%%%%%%%%%%%%%%%%%%%%%%%%%%%%%%%

These results are naturally understood 
if we draw the energy bands in the paramagnetic phase and the AF phase 
as in Figs.~\ref{fig:FS}(c) and \ref{fig:FS}(d), respectively. 
In Fig.~\ref{fig:FS}(c), the lower hybridized band of Hamiltonian (\ref{eq:PAM}) 
is occupied as illustrated by the solid line 
forming the ``large" Fermi surface whose volume contains f electrons, i.e., 
$\bar{n}_{\rm f}+\bar{n}_{\rm c}$. 
On the other hand, the energy band of conduction electrons is illustrated in Fig.~\ref{fig:FS}(d) 
by the gray dashed line whose occupied part for $\bar{n}_{\rm c}$ is represented by the solid line, 
clearly indicating the ``small'' Fermi surface, in sharp contrast to Fig.~\ref{fig:FS}(c). 
Once the AF order occurs, the lower and upper hybridized bands are folded 
as shown in Fig.~\ref{fig:FS}(d). 
Since the lower-hybridized band in the magnetic Brillouin zone is fully occupied, 
the Fermi surface emerging at the folded lower-hybridized band 
shows the area occupied for the filling 0.8 among $n=(1+0.8)/2$, 
giving rise to the same Fermi surface of the conduction band at $\bar{n}_{\rm c}=0.8$. 
Namely, the ``small" Fermi surface appears in the AF phase even though the c-f hybridization 
remains finite 
$\langle f^{\dagger}_{{\bf k}\sigma}c_{{\bf k}\sigma} \rangle\ne 0$. 
Thus, the mechanism of the coincidence of the Fermi surface in the AF phase and the dashed line shown in Fig.~\ref{fig:FS}(a) 
is naturally understood. 

We find that the total density of states 
$D(\mu)\equiv\sum_{{\bf k}\sigma}\delta(\mu-E_{{\bf k}\sigma})/(2N)$ 
at the Fermi level shows an enhancement toward $\varepsilon_{\rm f}=\varepsilon_{\rm f}^{\rm c}$ 
as shown in Fig.~\ref{fig:PD}(c). 
Here, $\mu$ is a chemical potential and $E_{{\bf k}\sigma}$ is the quasiparticle energy band 
with $\sigma$ spin. 
The reason why such a mass enhancement occurs at the AF-paramagnetic boundary is explained 
as follows: 
In the AF phase, 
as $\varepsilon_{\rm f}$ increases, the gap between the folded lower-hybridized-band and the original 
lower-hybridized-band increases (see Fig.~\ref{fig:FS}(d)). 
Then, the f-electron-dominant flat part of the folded band approaches the Fermi level, 
giving rise to the mass enhancement. 
In the paramagnetic phase, as $\varepsilon_{\rm f}$ decreases, $\bar{n}_{\rm f}$ increases to 
reach the $\bar{n}_{\rm f}=1$ state, i.e., to approach the Kondo regime. 
Hence, as $\varepsilon_{\rm f}$ approaches $\varepsilon_{\rm f}^{\rm c}$, mass enhancement 
occurs. 
Our result clearly shows that the mass enhancement is caused by the band effect including the 
local correlations of f electrons. 
This explains the experimental fact that the $\sqrt{A}/m^{*}=$const. scaling holds 
under pressure from $P=0$ to $P=P_{\rm c}$~\cite{Knebel2008}, where $A$ is the coefficient of 
the $T^{2}$ term in the low-temperature resistivity at $H=15$~T and $m^{*}$ is a cyclotron mass of 
the $\beta_{2}$ branch of the cylindrical Fermi surface observed for $H=12\sim 17$~T 
in CeRhIn$_5$~\cite{Shishido2005}. 
For $P>P_{\rm c}$, the dHvA signal of the $\beta_{2}$ branch was not detected~\cite{Shishido2005}, 
probably because of heavy effective mass larger than $m^{*}\approx 90m_{0}$ in CeCoIn$_5$. 
This is also consistent with our result shown in Fig.~\ref{fig:PD}(c). 

The Fermi wavenumber $k_{\rm F}$ defined by the crossing point of the line connecting 
${\bf k}=(0,0)$ and $(\pi,\pi)$, and the Fermi surface with minority spin at $h=0.005$ 
exhibits the abrupt change at the AF-paramagnetic boundary as shown in the inset of Fig.~\ref{fig:PD}(c). 
Thus, the drastic change of Fermi surfaces as well as the mass enhancement observed 
in the dHvA measurement~\cite{Shishido2005} is naturally understood by our theory. 
We note that the density of states 
at $\varepsilon_{\rm f}=-0.4$ is about 10 times larger than that of
conduction electrons at $\bar{n}_{\rm c}=0.8$, $D_{\rm c}(\mu)
\equiv\sum_{\bf k}\delta(\mu-\varepsilon_{\bf k})/N=0.092$, which is
also consistent with the enhanced $\gamma$ of CeRhIn$_5$ 
from that of LaRhIn$_5$ at $P=0$~\cite{Shishido2005}.

\section{Summary}
We have proposed a theory for resolving the outstanding puzzle 
on Fermi-surface change and transport anomalies in CeRhIn$_5$ 
in a unified way. 
The key mechanism has been shown to be the interplay of the AF order and Ce-valence fluctuations. 
Experimental tests to examine our proposal, in particular, direct observations of the Ce valence change 
under pressure and magnetic field near $P=P_{\rm c}$ are greatly desired.

\end{document}